\documentclass[prc,aps,onecolumn,showpacs,nofootinbib,superscriptaddress,preprint]{revtex4-1}
\usepackage{graphicx}
\usepackage{amssymb}
\usepackage{amsmath}

\allowdisplaybreaks

\begin{document}

\title{Amplitude reconstruction from complete experiments and
truncated partial-wave expansions}

\author{R. L. Workman}
\affiliation{Institute for Nuclear Studies and Department of Physics, The
George Washington University, Washington, DC 20052, USA}

\author{L. Tiator}
\affiliation{Institut f\"ur Kernphysik der Universit\"at Mainz,
Johann-Joachim-Becher-Weg 45, 55099 Mainz, Germany}

\author{Y. Wunderlich}
\affiliation{Helmholtz-Institut f\"ur Strahlen- und Kernphysik der
Universit\"at Bonn, Nu{\ss}allee 14-16, 53115 Bonn, Germany}

\author{M. D\"oring}
\affiliation{Institute for Nuclear Studies and Department of Physics, The
George Washington University, Washington, DC 20052, USA}

\affiliation{Thomas Jefferson National Accelerator Facility, 12000 Jefferson
Avenue, Newport News, VA 23606, USA}

\author{H. Haberzettl}
\affiliation{Institute for Nuclear Studies and Department of Physics, The
George Washington University, Washington, DC 20052, USA}

\date{\today}

\begin{abstract}

We compare the methods of amplitude reconstruction, for a complete experiment
and a truncated partial wave analysis, applied to the photoproduction of
pseudo-scalar mesons. The approach is pedagogical, showing in detail how the
amplitude reconstruction (observables measured at a single energy and angle)
is related to a truncated partial-wave analysis (observables measured at a
single energy and a number of angles).
\end{abstract}


\maketitle

\section{Introduction and Motivation}

A model-independent determination of amplitudes from experimental data is
mathematically possible, ignoring experimental errors, if one measures a
sufficient number of observables at a given energy and angle. This has been
done in nucleon-nucleon scattering~\cite{nn} and can be done~\cite{chiang,bds},
in principle, using pseudo-scalar meson photoproduction
data~\cite{vrancx,nys,sandorfi}.

The complete experiment analysis (CEA) determines helicity or transversity
amplitudes only up to an overall phase. This is a problem if one actually wants
partial-wave amplitudes, as the undetermined phase may be different at each
reconstructed energy and angle. In the analysis of pseudo-scalar
photoproduction data, we {\it do} require multipole amplitudes in order to
search for resonance content, and this has led to a renewed
interest~\cite{wunder,rw1} in the properties of a truncated partial-wave
analysis (TPWA), as has been described by Omelaenko~\cite{omel} and
Grushin~\cite{grushin}.

The number of required observables is different for the CEA and TPWA. The
reason for this is obscured by the fact that very different methods have been
used to derive the necessary conditions for a solution. Here, we have used
several methods to clarify the connections between the two approaches. The
first non-trivial example reveals many of these connections.

\section{Amplitudes used in pseudo-scalar meson photoproduction}

Before comparing the CEA and TPWA approaches, we review the notation used to
analyze pseudo-scalar photoproduction data. The multipoles and helicity
amplitudes are related by~\cite{Walker,vpi90}
\begin{subequations}
\begin{align}
H_1 &= {\frac{1}{\sqrt{2}}} \cos {\theta \over 2} \sin \theta \sum_{\ell =
1}^{\infty} \left[ E_{\ell +} - M_{\ell +} - E_{(\ell + 1) - }
-  M_{(\ell + 1) -} \right] \left( P_{\ell}^{''} - P_{\ell + 1}^{''} \right)~,
\\
H_2 &= {\frac{1}{\sqrt{2}}} \cos {\theta \over 2} \sum_{\ell = 0}^{\infty}
\left[ (\ell + 2)E_{\ell +} + \ell M_{\ell +} + \ell E_{(\ell + 1) - }
- (\ell +2) M_{(\ell + 1) -} \right] \left( P_{\ell}^{'} - P_{\ell + 1}^{'} \right)~,
\\
H_3 &= {\frac{1}{\sqrt{2}}} \sin {\theta \over 2} \sin \theta \sum_{\ell = 1}^{\infty}
\left[ (E_{\ell +} - M_{\ell +} +  E_{(\ell + 1) - }
+  M_{(\ell + 1) -} \right] \left( P_{\ell}^{''} + P_{\ell + 1}^{''} \right)~,
\\
H_4 &= {\frac{1}{\sqrt{2}}} \sin {\theta \over 2} \sum_{\ell = 0}^{\infty}
\left[ (\ell + 2)E_{\ell +} + \ell M_{\ell +} - \ell E_{(\ell + 1) - }
+ (\ell +2) M_{(\ell + 1) -} \right] \left( P_{\ell}^{'} + P_{\ell + 1}^{'} \right)~.
\end{align}
\end{subequations}
From these one can construct the transversity amplitudes~\cite{bds},
\begin{subequations}
\begin{align}
b_1 &= \frac{1}{2} \left[ \left( H_1 + H_4 \right) \; + \; i \; \left( H_2 - H_3 \right) \right]~,
 \\
b_2 &= \frac{1}{2} \left[ \left( H_1 + H_4 \right) \; - \; i \; \left( H_2 - H_3 \right) \right]~,
 \\
b_3 &= \frac{1}{2} \left[ \left( H_1 - H_4 \right) \; - \; i \; \left( H_2 + H_3 \right) \right]~,
 \\
b_4 &= \frac{1}{2} \left[ \left( H_1 - H_4 \right) \; + \; i \; \left( H_2 + H_3 \right) \right]~.
\end{align}
\end{subequations}
In Table~\ref{tab:obs1}, expressions for the observables of Type $S$ (cross
section and single-polarization), $BT$ (beam-target polarization), $BR$
(beam-recoil polarization), and $TR$ (target-recoil polarization) are given in
terms of both helicity and transversity amplitudes.

\begin{table*}
\begin{center}
\caption{\label{tab:obs1}Spin observables expressed in terms of helicity and
transversity amplitudes. Helicity amplitudes follow Walker~\cite{Walker} and
the SAID convention~\cite{vpi90}. The relations in Barker, Donnachie and
Storrow~\cite{Barker75} are adopted for observables with the replacement
$N\rightarrow H_2,\; S_1\rightarrow H_1,\; S_2\rightarrow H_4,\; D\rightarrow
H_3$. The transversity representation for $H$ corrects a typographical error in
Ref.~\cite{Barker75}. $I = (k/q)\,d\sigma / d\Omega$, $k$ and $q$ being the
photon and pion center-of-mass momenta. The checked observables, $\check{O}$,
are defined by $\check{O} = I\, O$. }
\bigskip
\begin{tabular}{|c|c|c|c|}
\hline
 Observable   &  Helicity                                             & Transversity  &                        Type  \\[-1.7ex]
              &  representation                                       & representation  &                            \\
\hline
          $I$      & $\frac{1}{2}$ ($|H_1|^2 + |H_2|^2 + |H_3|^2 + |H_4|^2$)  & $\frac{1}{2} (|b_1|^2+|b_2|^2+|b_3|^2+|b_4|^2$)   &        \\
      $\check{\Sigma}$      & Re$(H_1 H_4^* - H_2 H_3^*)$                           & $\frac{1}{2}(|b_1|^2+|b_2|^2-|b_3|^2-|b_4|^2)$   &  $S$   \\
           $\check{T}$      & Im$(H_1 H_2^* + H_3 H_4^*)$                           & $\frac{1}{2}(|b_1|^2-|b_2|^2-|b_3|^2+|b_4|^2$)   &        \\
           $\check{P}$      & $-$Im$(H_1 H_3^* + H_2 H_4^*)$                        & $\frac{1}{2}(|b_1|^2-|b_2|^2+|b_3|^2-|b_4|^2$)   &        \\
\hline
           $\check{G}$      & $-$Im$(H_1 H_4^* + H_2 H_3^*)$                        & ${\rm Im}(b_1 b_3^* + b_2 b_4^*)$    &        \\
           $\check{H}$      & $-$Im$(H_1 H_3^* - H_2 H_4^*)$                        & ${\rm Re}(b_1 b_3^* - b_2 b_4^*)$    &  $BT$  \\
           $\check{E}$      & $\frac{1}{2}(-|H_1|^2 + |H_2|^2 - |H_3|^2 + |H_4|^2)$ & $-{\rm Re}(b_1 b_3^* + b_2 b_4^*)$    &        \\
           $\check{F}$      & Re$(H_1 H_2^* + H_3 H_4^*)$                           & ${\rm Im}(b_1 b_3^* - b_2 b_4^*)$    &        \\
\hline
      $\check{O}_{x} $      & $-$Im$(H_1 H_2^* - H_3 H_4^*)$                        & $-{\rm Re}(b_1 b_4^* - b_2 b_3^*)$  &       \\
      $\check{O}_{z} $      & Im$(H_1 H_4^* - H_2 H_3^*)$                           & $-{\rm Im}(b_1 b_4^* + b_2 b_3^*)$  &  $BR$ \\
      $\check{C}_{x} $      & $-$Re$(H_1 H_3^* + H_2 H_4^*)$                        & ${\rm Im}(b_1 b_4^* - b_2 b_3^*)$   &       \\
      $\check{C}_{z} $      & $\frac{1}{2}(-|H_1|^2 - |H_2|^2 + |H_3|^2 + |H_4|^2)$ & $-{\rm Re}(b_1 b_4^* + b_2 b_3^*)$   &       \\
\hline
      $\check{T}_{x} $     & Re$(H_1 H_4^* + H_2 H_3^*)$                           & ${\rm Re}(b_1 b_2^* - b_3 b_4^*)$  &       \\
      $\check{T}_{z} $     & Re$(H_1 H_2^* - H_3 H_4^*)$                           & ${\rm Im}(b_1 b_2^* - b_3 b_4^*)$  &  $TR$ \\
      $\check{L}_{x} $     & $-$Re$(H_1 H_3^* - H_2 H_4^*)$                        & ${\rm Im}(b_1 b_2^* + b_3 b_4^*)$  &       \\
      $\check{L}_{z} $     & $\frac{1}{2}(|H_1|^2 - |H_2|^2 - |H_3|^2 + |H_4|^2)$  & ${\rm Re}(b_1 b_2^* + b_3 b_4^*)$  &       \\
\hline
\end{tabular}
\end{center}
\end{table*}

Transversity amplitudes often simplify the discussion of amplitude
reconstruction, as the type-$S$ observables determine their moduli. Another
simplification is the property
\begin{equation}
b_2(\theta) = -b_1(- \theta) \quad {\rm and} \quad
b_4(\theta) = -b_3(- \theta)~, \label{eq:TraAmplsSymmetryRelations}
\end{equation}
which allows one to parameterize only two of the four transversity amplitudes.
The form introduced by Omelaenko,
\begin{subequations}\label{eq:BallOriginalOmelDecomposition}
\begin{align}
b_1 &= c\,a_{2L} {{e^{i\theta /2} }\over {(1+x^2)^L}} \prod_{i=1}^{2L} (x-\alpha_i)~,
\label{eq:B1OriginalOmelDecomposition}
\\
b_3 &= -c\,a_{2L} {{e^{i\theta /2} }\over {(1+x^2)^L}} \prod_{i=1}^{2L} (x-\beta_i)~,
\label{eq:B3OriginalOmelDecomposition}
\end{align}
\end{subequations}
with $x=\tan ( \theta /2 )$ and $L$ being the upper limit for $\ell$, is
convenient for a truncated partial wave analysis, as the ambiguities can be
linked to the conjugation of the complex roots of the above relations, with a
constraint
\begin{equation}
\prod_{i=1}^{2L} \alpha_i \; = \; \prod_{i=1}^{2L} \beta_i~.
\label{eq:OriginalOmelConstraint}
\end{equation}
The quantities $a_{2L}$ and $c$ above will be clarified in an explicit example described in
Sec.~III below.

In a complete experiment analysis (CEA), one attempts to determine the
transversity or helicity amplitudes, based on the relations in Table I, at a
particular energy and angle. Barker, Donnachie and Storrow~\cite{bds} (BDS)
showed how this could be done with 9 well-chosen observables. For example, the
case of $I$, $\check{P}$, $\check{\Sigma}$, $\check{T}$,
$\check{E}$, $\check{F}$, $\check{G}$, $\check{L}_x$, and $\check{L}_z$ was
worked out explicitly in Ref.~\cite{bds}. More recently, a counter-example to
this scheme was noticed in Ref.~\cite{KW}, which led to the finding, by Chiang
and Tabakin~\cite{chiang}, that it was possible to perform a CEA with one
measurement less. In the case presented by BDS, Chiang and Tabakin
demonstrated a solution with only $I$, $\check{P}$,
$\check{\Sigma}$, $\check{T}$, $\check{F}$, $\check{G}$, $\check{T}_x$, and
$\check{L}_x$ being required.

In a truncated partial-wave analysis (TPWA), the multipole expansion of helicity or transversity
amplitudes is cut off at some upper limit $L$. Here one finds the amplitudes, for all angles, at a
particular energy. Omelaenko showed how this can be done, eliminating the root-conjugation
ambiguities associated with the transversity amplitudes in
Eqs.~(\ref{eq:BallOriginalOmelDecomposition}), using an $L$-dependent number of angular
measurements of five observables, such as $I$, $\check{P}$, $\check{\Sigma}$, $\check{T}$, and
$\check{F}$. As the methods of proof are very different, in the CEA and TPWA problems, it is not
obvious how these results can be compared. In the following, we compare the CEA and TPWA results in
such a way that the differences can be more easily understood.

\section{Amplitude Reconstruction}

\subsection{\boldmath Trivial case: $L = 0$}

It is instructive to compare methods starting with the trivial $\ell = 0$ case of a
single $E_{0+}$ multipole and build up to the case studied by Omelaenko~\cite{omel}
including the $E_{0+}$, $M_{1-}$, $E_{1+}$, and $M_{1+}$ multipoles.
If only one complex amplitude ($E_{0+}$) is included, from Eq.~(1) we see that there are
2 non-zero helicity amplitudes ($H_2$ and $H_4$) which are related by a real factor.
Here, we may simply measure the cross section at a single angle. While this gives only
one real number, and the amplitudes are complex, the fact that observables involve only
bilinear products of amplitudes, i.e. terms of the form $A^*B$, prevents the measurement of any
overall phase associated with the amplitudes.  This solves both the CEA and TPWA with the
same experimental input.

\subsection{\boldmath Simplest non-trivial case: $J= 1/2$}

The first non-trivial case includes the $E_{0+}$ and $M_{1-}$ multipoles, i.e.
partial waves with $J=1/2$. This combination again produces 2 non-zero helicity
amplitudes ($H_2$ and $H_4$). In this case, however, the amplitudes are
independent. The corresponding transversity amplitudes are given by%
\footnote{The corresponding expression in Ref.~\cite{omel} differs by an
   overall
   phase ($-i$) and a factor $\sqrt{2}c$ which incorporates the kinematic factor
   of Table I, here converting $d\sigma /dt$ to $I$, into the definition
   of the transversity amplitudes.}
\begin{equation}
  b_1 = {i\over {\sqrt{2}}} \left( -e^{i\theta /2} E_{0+} + e^{-i\theta /2} M_{1-} \right)~,
   \label{eq:B1JOneHalf}
\end{equation}
with $b_3 = -b_1$, and with $(b_2, b_4)$ given by Eq.~(3). Below, in Table II,
we give the observables both in terms of the helicity/transversity amplitudes
(CEA) and the 2 included multipoles (TPWA).


\begin{table*}
\begin{center}
\caption{\label{tab:obs2} Spin observables in terms of helicity amplitudes for
a CEA and multipole amplitudes for a TPWA with $J=1/2$. Here we have used
$b_3= -b_1$ and $b_4=-b_2$ }
\begin{tabular}{|c|c|c|c|c|}
\hline
 Observable   &  CEA (Helicity)  & CEA (Transversity)        & TPWA  &                        Type  \\
\hline
         $I$      & $\frac{1}{2}\left(|H_2|^2 + |H_4|^2\right)$ & $|b_1|^2 + |b_2|^2$     & $\left(|E_{0+}|^2 + |M_{1-}|^2\right) - 2\cos\theta\,{\rm Re}\left(E_{0+}M_{1-}^*\right)$   &        \\
      $\check{\Sigma}$      & 0  & 0                                    & 0   &  $S$   \\
           $\check{T}$      & 0  & 0                         & 0   &        \\
           $\check{P}$      & $-$Im$\left(H_2 H_4^*\right)$ & $|b_1|^2 - |b_2|^2$                  & $2\sin \theta\,{\rm Im}\left(E_{0+}M_{1-}^*\right)$   &        \\
\hline
           $\check{G}$      & 0  & 0                  & 0    &        \\
           $\check{H}$      & $$Im$\left(H_2 H_4^*\right)$ & $-|b_1|^2 + |b_2|^2$                   & $- 2\sin \theta\,{\rm Im}\left(E_{0+}M_{1-}^*\right)$    &  $BT$  \\
           $\check{E}$      & $\frac{1}{2}\left(|H_2|^2 + |H_4|^2\right)$ & $|b_1|^2 + |b_2|^2$    & $\left(|E_{0+}|^2 + |M_{1-}|^2\right) - 2\cos\theta\,{\rm Re}\left(E_{0+}M_{1-}^*\right)$    &        \\
           $\check{F}$      & 0    & 0                    & 0    &        \\
\hline
      $\check{O}_{x} $      & 0   & 0                     & 0  &       \\
      $\check{O}_{z} $      & 0   & 0                        & 0  &  $BR$ \\
      $\check{C}_{x} $      & $-$Re$\left(H_2 H_4^*\right)$ & $-2$Im$b_1 b_2^*$          & $\sin \theta\,\left(|E_{0+}|^2 - |M_{1-}|^2\right)$   &       \\
      $\check{C}_{z} $      & $\frac{1}{2}\left(-|H_2|^2 + |H_4|^2\right)$ & 2Re$b_1 b_2^*$ & $2{\rm Re}\left(E_{0+} M_{1-}^*\right) - \cos \theta\,\left(|E_{0+}|^2 + |M_{1-}^*|^2\right)$   &       \\
\hline
      $\check{T}_{x} $     & 0 & 0                          & 0  &       \\
      $\check{T}_{z} $     & 0 & 0                          & 0  &  $TR$ \\
      $\check{L}_{x} $     & Re$\left(H_2 H_4^*\right)$ & 2Im$b_1 b_2^*$           & $-\sin \theta\,\left(|E_{0+}|^2 - |M_{1-}|^2\right)$  &       \\
      $\check{L}_{z} $     & $\frac{1}{2}\left(-|H_2|^2 + |H_4|^2\right)$  & 2Re$b_1 b_2^*$ & $2{\rm Re}\left(E_{0+} M_{1-}^*\right) - \cos \theta\,\left(|E_{0+}|^2 + |M_{1-}^*|^2\right)$  &       \\
\hline
\end{tabular}
\end{center}
\end{table*}

Here the CEA requires 4 measurements at a given energy and angle. For example,
$I$, $\check{P}$, plus either the Beam-Recoil sets
($\check{C}_x$ and $\check{C}_z$) or the Target-Recoil ($\check{L}_x$ and
$\check{L}_z$). The TPWA requires one fewer observable, a possible choice being
$I$, $\check{P}$, and $\check{C}_x$ or $\check{L}_x$,
compensated by a second angular measurement of the cross section.

For both the CEA and TPWA, closed expressions for the solution of the inverse problem can be
obtained in this special case $J = 1/2$. It is instructive to work them out explicitly. For the
quantities $I$, $\check{P}$ and $\check{C}_{x}$ in the TPWA, it is possible to parametrize the
angular dependence given in Table \ref{tab:obs2} as
\begin{equation}
 I = \sigma_{0} + \sigma_{1}\, \cos \theta~,\quad
  \check{P} = P_0\,\sin \theta ~, \quad
  \check{C}_{x} = C_{x0}\,\sin \theta ~,
   \label{eq:AngDistsAobs}
\end{equation}
where each coefficient carries the energy dependence of the multipoles. It is
clear that to extract values for $\sigma_{0}$, $\sigma_{1}$, $P_{0}$ and
${C}_{x0}$, both spin asymmetries and the cross section are needed at the same
angle, with an additional angular measurement required for the cross section.

Having obtained the four coefficients, the zeroth order quantities of $I$ and
$\check{C}_{x}$ can be directly solved for the moduli of the multipoles (cf.\
Table \ref{tab:obs2}),
\begin{equation}
 \left| E_{0+} \right| = \sqrt{\frac{\sigma_{0} + C_{x0}}{2}}~,
 \quad
 \left| M_{1-} \right| =  \sqrt{\frac{\sigma_{0} - {C}_{x0}}{2}}~.
 \label{eq:ModuliOfJOneHalfMultipoles}
\end{equation}
The relative phase $\phi_{E,M} \equiv \phi_{E} - \phi_{M}$ between the
multipoles $E_{0+}$ and $M_{1-}$ is obtainable via the remaining two
coefficients, both containing information on the real and imaginary parts of
the bilinear product $E_{0+} M_{1-}^{\ast}$. The additional angular measurement
for the cross section fixes the real part,
\begin{equation}
 \mathrm{Re} \left( E_{0+} M_{1-}^{\ast} \right)
 = \left| E_{0+} \right| \left| M_{1-} \right|  \mathrm{Re} \left( e^{i \phi_{E,M}} \right) = - \frac{1}{2} \sigma_{1}~,
 \label{eq:ReBilMultProduct}
\end{equation}
while the imaginary part can be extracted from the single measurement of
$\check{P}$,
\begin{equation}
 \mathrm{Im} \left( E_{0+} M_{1-}^{\ast} \right) = \left| E_{0+} \right| \left| M_{1-} \right|  \mathrm{Im} \left( e^{i \phi_{E,M}} \right) = \frac{1}{2} {P}_{0}~.
  \label{eq:ImBilMultProduct}
\end{equation}
Together, these define the exponential of the relative phase, provided that
none of the moduli vanish,
\begin{equation}
  e^{i \phi_{E,M}} = \frac{- \sigma_{1} + i {P}_{0}}{\sqrt{\sigma_{0} + {C}_{x0}} \sqrt{\sigma_{0} - {C}_{x0}}}~.
  \label{eq:PhaseBilMultProductExtracted}
\end{equation}
This function can be inverted uniquely on the interval $\left[ 0, 2 \pi \right)$. Therefore, no quadrant ambiguity remains.
The multipoles have been extracted up to an overall phase.

The CEA proceeds in a mathematically exactly analogous way. The observables
$I$, $\check{P}$, $\check{C}_{x}$ and $\check{C}_{z}$ yield the moduli
and relative phase $\phi_{1,2} \equiv \phi_{b_{1}} - \phi_{b_{2}}$ of the
transversity amplitudes using exactly the same calculation (cf.\ Table
\ref{tab:obs2})
\begin{align}
 \left| b_{1} \right| =& \sqrt{\frac{I + \check{P}}{2}}~,
 \quad\left| b_{2} \right| = \sqrt{\frac{I - \check{P}}{2}}~,
 \label{eq:ModuliOfTrAmplitudes} \\
  & e^{i \phi_{1,2}}  = \frac{\check{C}_{z} - i \check{C}_{x}}{\sqrt{I
   + \check{P}} \sqrt{I - \check{P}} }~.
   \label{eq:RelPhaseOfTrAmplitudes}
\end{align}
A crucial difference, however, lies in the kinematical regions over which the
CEA and TPWA operate. For a fixed energy, the CEA extracts amplitudes from
observables at exactly the same angle and it is completely blind to what may
happen at neighbouring angles. The TPWA uses
the angular distributions of the observables which, in the present
case of $I(\theta)$, is linear in $\cos \theta$.
One seemingly obtains a reduction from $4$
to $3$ observables, but this is bought at the price of having to measure
angular distributions which become, for the higher truncation orders,
increasingly complicated.

The difference in the nature of these analyses also becomes obvious in
considering the end results they yield. The CEA returns transversity amplitudes
only at a single angle, up to an energy- and angle-dependent overall phase, cf.
Eqs.~(\ref{eq:ModuliOfTrAmplitudes}) and (\ref{eq:RelPhaseOfTrAmplitudes}).
However, from the result of the TPWA, the moduli
Eq.~(\ref{eq:ModuliOfJOneHalfMultipoles}) and relative phase
Eq.~(\ref{eq:PhaseBilMultProductExtracted}) of multipoles, it is possible to
infer transversity amplitudes at all angles, this time up to an
energy-dependent phase.

\subsection{\boldmath Unique features of the $J=1/2$ case}

It is useful to compare the special case of $J=1/2$ to more general results for
the CEA and TPWA in Refs.~\cite{chiang} and \cite{omel}. In Ref.~\cite{chiang},
a complete set of 8 experiments, explicitly derived and compared to the
corresponding BDS case (requiring 9 experiments) is: ($I$,
$\check{\Sigma}$, $\check{P}$, $\check{T}$, $\check{G}$, $\check{F}$,
$\check{L}_x$, $\check{T}_x$). Here, with a truncation to $J=1/2$, this set
becomes ($I$, 0, $\check{P}$, 0, 0, 0, $\check{L}_x$, 0), which
does not contain sufficient information, as can be seen directly from Table~II.
However, the older BDS set, which exchanges $\check{T}_x$ for $\check{E}$ and
$\check{L}_z$, {\it does} constitute a complete experiment. This failure of a
set of 8 experiments is due to the number of zero quantities in Table~II. The
effect can be seen in constraint equation (4.10) employed in the derivation of
Ref.~\cite{chiang}. Many Fierz identities listed in Ref.~\cite{chiang}
similarly revert to zero-equals-zero relations in this special case.

Similarly, the TPWA conditions for a complete set~\cite{omel}, derived for a
case including the $E_{0+}$, $M_{1-}$, $E_{1+}$ and $M_{1+}$ multipoles, do not
directly reduce to the result given here if the $E_{1+}$ and $M_{1+}$
multipoles are simply set to zero. In Refs.~\cite{omel,wunder}, a complete set
is given as ($I$, $\check{P}$, $\check{\Sigma}$, $\check{T}$,
$\check{G}$), which again is insufficient in this special case.

To understand how a truncation to $J=1/2$ changes the result, it is instructive
to repeat Omelaenko's analysis \cite{omel}, which led to the general
parametrizations of Eq.~(\ref{eq:B1OriginalOmelDecomposition}) and
Eq.~(\ref{eq:B3OriginalOmelDecomposition}), under the constraint in Eq.
(\ref{eq:OriginalOmelConstraint}), for all $L \geq 1$.

Expressing $\cos \theta$ and $\sin \theta$ in terms of $x=\tan ( \theta /2 )$,
one can write
\begin{equation}
 e^{i \theta} = \frac{\left( 1 + i x \right)^{2}}{1 + x^{2}}~.
  \label{eq:ExponentialInTermsOfTanThetaHalf}
\end{equation}
Starting from the expression for $b_{1}$ in terms of multipoles given in
Eq.~(\ref{eq:B1JOneHalf}), we have
\begin{align}
 b_1 &= {i\over {\sqrt{2}}} \left( -e^{i\theta /2} E_{0+} + e^{-i\theta /2} M_{1-} \right) \nonumber \\
 &= \frac{i e^{i\theta /2}}{\sqrt{2}} \left( - E_{0+} + \frac{1 - x^{2} - 2 i x}{(1 + x^{2})} M_{1-} \right) \nonumber \\
 &= \frac{-i}{\sqrt{2}} \frac{e^{i\theta /2}}{(1 + x^{2})} \left( E_{0+} + M_{1-} \right) \left( x^{2} + \frac{2 i M_{1-}}{E_{0+} + M_{1-}} x + \frac{E_{0+} - M_{1-}}{E_{0+} + M_{1-}} \right) \nonumber \\
 &\equiv \frac{-i}{\sqrt{2}} \frac{e^{i\theta /2}}{(1 + x^{2})}  a_{2} \left( x^{2} + \hat{a}_{1} x + \hat{a}_{0} \right)~.
  \label{eq:deriveB1Polynomial}
\end{align}
Note that the coefficients $a_{2}$, $\hat{a}_{1}$ and $\hat{a}_{0}$, defining
the amplitude in the last step, are fully equivalent to the multipoles.
Decomposing the polynomial into a product of linear factors defined by two
complex roots $\alpha_{1}$ and $\alpha_{2}$, the Omelaenko decomposition of the
amplitude $b_{1}$ is obtained as
\begin{equation}
 b_1 = \frac{-i}{\sqrt{2}}\, a_{2} e^{i\theta /2} \,\frac{\prod_{i=1}^{2} \left( x - \alpha_{i} \right)}{1 + x^{2}}  \mathrm{.} \label{eq:B1OmelaenkoDecomposition}
\end{equation}
The expression for the only remaining non-redundant amplitude, $b_{2}$, for $J = 1/2$,
is obtained by invoking the symmetry in Eq. (\ref{eq:TraAmplsSymmetryRelations}),
\begin{equation}
 b_2 (\theta) = - b_{1} (- \theta) = \frac{i}{\sqrt{2}} a_{2} e^{- i \theta /2} \frac{\prod_{i=1}^{2} \left( x + \alpha_{i} \right)}{1 + x^{2}}  \mathrm{.} \label{eq:B2OmelaenkoDecomposition}
\end{equation}
Therefore, for $J = 1 / 2$ there are only two $\alpha$-roots, no $\beta$-roots and the constraint Eq. (\ref{eq:OriginalOmelConstraint})
no longer appears.

In view of the already obtained results
Eq.~(\ref{eq:ModuliOfJOneHalfMultipoles}) and
Eq.~(\ref{eq:PhaseBilMultProductExtracted}), the observable $\check{C}_{x}$
will have to be tested for its response to discrete ambiguity transformations.
The full Omelaenko decomposition of this observable becomes
\begin{equation}
 \check{C}_{x} = - 2 \mathrm{Im} b_{1} b_{2}^{\ast} = \frac{ \left| a_{2} \right|^{2} }{(1 + x^{2})^{2}}\,
  \mathrm{Im} \left[ e^{i \theta} \prod_{i=1}^{2} \left(x - \alpha_{i}\right) \left(x + \alpha_{i}^{\ast} \right) \right]~.
\label{eq:CxInTermsOfOmelaenkoRoots}
\end{equation}
The decompositions of amplitudes $b_{i}$ in terms of roots $\alpha_{1}$ and
$\alpha_{2}$ given by Eqs.~(\ref{eq:B1OmelaenkoDecomposition}) and
(\ref{eq:B2OmelaenkoDecomposition}) facilitate a study of the discrete
ambiguities of the quantities $I$ and $\check{P}$ (as well as
$\check{E}$ and $\check{H}$), since they are just linear combinations of the
squared moduli $\left| b_{1} \right|^{2}$ and $\left| b_{2} \right|^{2}$ (see
Table~\ref{tab:obs2}). The ambiguities are obtained by the complex conjugation
of subsets of roots, as stated below
Eqs.~(\ref{eq:BallOriginalOmelDecomposition}).

Note that the multipoles $E_{0+}$ and $M_{1-}$, with an undetermined overall
phase that can be arbitrarily fixed, correspond to $3$ real numbers. For the
variables $\left(a_{2}, \alpha_{1}, \alpha_{2}\right)$ of the Omelaenko
decomposition, where the phase of $a_2$ cannot be determined, one counts $5$
real degrees of freedom. The general constraint
equation~(\ref{eq:OriginalOmelConstraint}), which is true for an expansion in
$\ell$ for all $L \geq 1$, is missing here. So, there must be another way in
which the effective number of real degrees of freedom is reduced from $5$ to
$3$.

One can learn more by considering the equations which relate the Omelaenko roots $\left(\alpha_{1}, \alpha_{2}\right)$ to the multipoles $\left(E_{0+}, M_{1-}\right)$. Utilizing the notation of
Eqs. (\ref{eq:deriveB1Polynomial}) and (\ref{eq:B1OmelaenkoDecomposition}), we have
\begin{equation}
 \hat{a}_{1} \equiv - \left( \alpha_{1} + \alpha_{2} \right) = \frac{2 i M_{1-}}{E_{0+} + M_{1-}}~,
 \quad \hat{a}_{0} \equiv \alpha_{1} \alpha_{2} = \frac{E_{0+} - M_{1-}}{E_{0+} + M_{1-}}  \mathrm{.} \label{eq:EqsGiveRootsInTermsOfMultipoles}
\end{equation}
These relations lead to a quadratic equation with two solutions given by the roots
\begin{equation}
 \alpha_{1}^{(\mathrm{I})} = i \, \frac{E_{0+} - M_{1-}}{E_{0+} + M_{1-}}~,
 \quad
 \alpha_{2}^{(\mathrm{I})} = -i
 \quad \text{and}
 \quad \alpha_{1}^{(\mathrm{II})} = -i~,
 \quad
 \alpha_{2}^{(\mathrm{II})} = i \, \frac{E_{0+} - M_{1-}}{E_{0+} + M_{1-}}~.
 \label{eq:JOneHalfRootSolutionCases}
\end{equation}
Both solutions remove the overcounting mentioned above. Two real degrees of
freedom are always removed since one of the roots has a fixed value. Only one
of the two roots depends on the multipoles.

Solutions I and II of (\ref{eq:JOneHalfRootSolutionCases}) are not distinct, as
both are equivalent by a simple re-labelling of the roots. Taking solution I,
for which $\alpha_{2}$ is fixed to $-i$, only one discrete ambiguity remains in
the Omelaenko formulation for $J=1/2$, represented by the transformation
$\alpha_1\to\alpha_1^*$,

Using solution I, the full Omelaenko decomposition of $\check{C}_{x}$,
Eq.~(\ref{eq:CxInTermsOfOmelaenkoRoots}), simplifies significantly. Again
writing the exponential $e^{i \theta}$ in terms of $x = \tan(\theta/2)$, see
Eq.~(\ref{eq:ExponentialInTermsOfTanThetaHalf}), we have the identity
\begin{equation}
 e^{i \theta}\, \left( x - \alpha_{2} \right) \left( x + \alpha_{2}^{\ast} \right)
 = \frac{\left( 1 + i x \right)^{2}}{1 + x^{2}} \, \left( x + i \right)^{2}
 = - \left( 1 + x^{2} \right)~.
  \label{eq:EqWhichHelpsToReduceCx}
\end{equation}
The expression for $\check{C}_{x}$, in terms of the only non-redundant Omelaenko root, $\alpha_{1}$, then becomes
\begin{equation}
 \check{C}_{x} = \frac{ - \left| a_{2} \right|^{2} }{1 + x^{2}} \mathrm{Im} \left[ \left(x - \alpha_{1}\right) \left(x + \alpha_{1}^{\ast} \right) \right]~.
  \label{eq:CxInTermsOfOmelaenkoRootsSimplified}
\end{equation}

For the discrete symmetry, $\alpha_1\to\alpha_1^*$, we see that expression
(\ref{eq:CxInTermsOfOmelaenkoRootsSimplified}) changes sign, $\check{C}_{x}
\rightarrow  - \check{C}_{x} $, once the ambiguity transformation is applied.
Furthermore, $\check{C}_{x}$ generally only remains invariant at the angles
$\theta = 0$ and $\theta = \pi$, where it vanishes by definition (see Table
\ref{tab:obs2}).

Another interesting special case is found if one requires the transformation
$\alpha_1\to\alpha_1^*$ to produce no ambiguity, which can only be fulfilled
for a real root. Once this condition is evaluated for the explicit form of
$\alpha_{1}$ in terms of multipoles, given in
Eq.~(\ref{eq:JOneHalfRootSolutionCases}), one finds that $\alpha_{1}^{\ast} =
\alpha_{1}$ is equivalent to $\left| E_{0+} \right| = \left| M_{1-} \right|$.

The Omelaenko decomposition of $\check{C}_{x}$, as well as the explicit form of
this quantity written in terms of multipoles (see Table \ref{tab:obs2}), shows
that in this particular case $\check{C}_{x}$ vanishes for all angles. Here,
while the sign information associated with $\check{C}_{x}$ may be missing, it
is not required, as the discrete symmetry, which is resolved precisely by this
sign, no longer exists. Also, Eqs.~(\ref{eq:ModuliOfJOneHalfMultipoles}) and
(\ref{eq:PhaseBilMultProductExtracted}) imply that in this special case, i.e.
$\left| E_{0+} \right| = \left| M_{1-} \right|$ or equivalently ${C}_{x0} = 0$,
the moduli of both multipoles, as well as the relative phase $\phi_{E,M}$, are
determined by $I$ and $\check{P}$ alone. This case is, however, the only
situation where a solution of the inverse problem is possible with just $2$
observables.

In summary, both the explicit inversion of the TPWA,
Eqs.~(\ref{eq:ModuliOfJOneHalfMultipoles}) and
(\ref{eq:PhaseBilMultProductExtracted}), and the study of the discrete
ambiguities, according to Omelaenko's method, yield consistent results for $J =
1 / 2$. This has been exemplified by the solvability of the example set
$I$, $\check{P}$ and $\check{C}_{x}$.
The case $J = 1 / 2$ is special as it allows all three analyses, the CEA,
TPWA and ambiguity study, to be performed using simple algebra. For the
higher orders $L \geq 1$, Chiang and Tabakin \cite{chiang} have published a
solution for the CEA which holds apart from the special case discussed above.

An algebraic inversion of the TPWA, i.e.\ the extraction of the bilinear
products of multipoles by an effective linearization of the problem, followed
by a simple evaluation of moduli and relative phases, does not appear to be
possible for $L \geq 1$. The only principle that carries through to the higher
orders is the study of discrete ambiguities \cite{omel, wunder}, using the
expressions in Eqs. (\ref{eq:B1OriginalOmelDecomposition}),
(\ref{eq:B3OriginalOmelDecomposition}) and (\ref{eq:OriginalOmelConstraint}).

In this way, complete sets of observables can still be proposed. However, the actual completeness
of such sets should, in any case, be checked by a full solution of the inverse problem which, for
the higher truncation orders, can only be done numerically.

\subsection{Counting Observables}

In examining the $J=1/2$ case, it was found that a formal solution was possible
with only $\check{P}$, $\check{C}_{x}$, and $\check{C}_{z}$
(3 rather than 4 quantities), measured at one angle, if one
used the overall phase freedom to make one amplitude real and positive. This
result could be understood by refining how the counting of observables is done.
If a measurement, done with a fixed beam, target and detector setup, produces
an `observable', then the measurement of a polarization asymmetry (spin up
versus spin down) is actually two observables. These two measurements can then
be combined to form both the asymmetry and the cross section. Once the cross
section is known, a second asymmetry can, in principle, be determined from only
one of the two possible (such as spins parallel versus anti-parallel)
measurements. Thus, the set ($\check{P}$, $\check{C}_{x}$, $\check{C}_{z}$)
requires 2+1+1=4
measurements, compared to the set
($I$, $\check{P}$, $\check{C}_{x}$, $\check{C}_{z}$),
requiring 1+1+1+1=4 measurements.

\section{\boldmath Comparing CEA and TPWA beyond $J=1/2$}

In Table III, the examples discussed in detail above are generalized to higher
angular-momentum cutoffs. The examples with one, two, and three multipoles show
that in the CEA and TPWA approaches, the number of measurements is the same. In
cases where a TPWA is possible with all measurements at a single energy and
angle, the results are directly related. Note that in the case of 3 multipoles,
only 3 of the helicity/transversity amplitudes are independent. This is also
true for the standard set of 4 multipoles ($E_{0+}$, $M_{1-}$, $E_{1+}$,
$M_{1+}$) as can be most easily seen if, instead, one writes out the CGLN
amplitudes,
\begin{subequations}
\label{eq:CGLN}
\begin{align}
F_1(\theta) &= E_{0+}+3\,(M_{1+}+E_{1+})\,\mbox{cos}\,\theta~,\\
F_2(\theta) &= 2\,M_{1+}+M_{1-}~,\\
F_3(\theta) &= 3\,(E_{1+}-M_{1+})~, \\
F_4(\theta) &= 0  ~.
\end{align}
\end{subequations}
With $F_4=0$, only 3 independent amplitudes can be extracted in a CEA.
Consequently, also only 3 linear combinations of multipoles can be obtained in
an experiment at a single angle.

Extending the expansion of observables, given in Eq.~(8), to higher
orders in $\cos \theta$ up to the highest powers for a given $L$, we have
\begin{subequations}
\begin{align}
I &= \sigma_0 + \sigma_1\,\cos\,\theta + \sigma_2\,\cos^2\theta + \cdots + \sigma_{2L}\,\cos^{2L}\theta\,,\\
\check{\Sigma} &= \sin^2\,\theta\left( \Sigma_0 + \cdots + \Sigma_{2L-2}\,\cos^{2L-2}\theta\right)\,,  \\
\check{T} &= \sin\,\theta\left( T_0 + T_1\,\cos\theta + \cdots + T_{2L-1}\,\cos^{2L-1}\theta\right)\,, \\
\check{P} &= \sin\,\theta\left( P_0 + P_1\,\cos\theta + \cdots +
P_{2L-1}\,\cos^{2L-1}\theta\right)\,.
\end{align}
\end{subequations}
The remaining double-polarization observables, ($\check{E}$, $\check{C}_{x}$, $\check{O}_{x}$,
$\check{T}_{z}$, $\check{L}_{x}$) behave like $I$, ($\check{F}$, $\check{H}$,
$\check{O}_{z}$, $\check{T}_{x}$) like $\check{T}$, $\check{G}$ like $\check{\Sigma}$, while
$\check{C}_{z}$ and $\check{L}_{z}$ exhibit the highest powers up to $\cos^{2L+1}\theta$.

\begin{table}[t!]
\caption{Examples of measurements at a single energy for CEA and TPWA. The number of different
measurements (n), different observables (m) and different angles (k) needed for a complete analysis
are given as $n(m)k$. Entries with a $\dagger$ do not allow the comparison CEA $\leftrightarrow$
TPWA. For cases with only one angle, the CEA and TPWA are equivalent. The number of necessary
distinct angular measurements is given in brackets.
\\}\label{tab3}
\begin{tabular}{|c|l|c|c|l|}
\hline Set & Included Partial Waves & CEA & TPWA & Complete Sets for TPWA\\
\hline
1 &$L=0\; (E_{0+})$ & 1(1) & 1(1)1 & $I[1]$\\
\hline
2 &$J=1/2\; (E_{0+},M_{1-})$ & 4(4) & 4(4)1 & $I[1]\,,\check{P}[1]\,,\check{C}_{x}[1]\,,\check{C}_{z}[1]$\\
  &                          &      & 4(3)2 & $I[2]\,,\check{P}[1]\,,\check{C}_{x}[1]$\\
 \hline
3 &$L=0,1\; (E_{0+},M_{1-},E_{1+})$ & 6(6) & 6(6)1 & $I[1]\,,\check{\Sigma}[1]\,,\check{T}[1]\,,\check{P}[1]\,,\check{F}[1]\,,\check{G}[1]$\\
   &                                &      & 6(4)2 & $I[2]\,,\check{\Sigma}[1]\,,\check{T}[2]\,,\check{P}[1]$\\
   &                                &      & 6(3)3 & $I[3]\,,\check{\Sigma}[1]\,,\check{T}[2]$\\
 \hline
4 &$L=0,1\; (E_{0+},M_{1-},E_{1+},M_{1+})$ & $\dagger$  &    &  TPWA at 1 angle not possible\\
  & full set of 4 $S,P$ wave multipoles      &   & 8(5)2 & $I[2]\,,\check{\Sigma}[1]\,,\check{T}[2]\,,\check{P}[2]\,,\check{F}[1]$\\
  &                                          &   & 8(4)3 & $I[3]\,,\check{\Sigma}[1]\,,\check{F}[2]\,,\check{H}[2]$\\
 \hline
5 &$L=0,1,2\; (E_{0+},M_{1-},E_{1+},E_{2-})$ & 8(8) & 8(8)1 & $I[1]\,,\check{\Sigma}[1]\,,\check{T}[1]\,,\check{P}[1]\,,\check{F}[1]\,,\check{G}[1]\,,\check{C}_{x}[1]\,,\check{O}_{x}[1]$\\
   &                                          &   & 8(4)2 & $I[2]\,,\check{\Sigma}[2]\,,\check{T}[2]\,,\check{P}[2]$\\
   &                                          &   & 8(3)3 & $I[3]\,,\check{\Sigma}[2]\,,\check{T}[3]$\\
 \hline
6 &$J\leq3/2\; (E_{0+},M_{1-},E_{1+},M_{1+},E_{2-},M_{2-})$  & $\dagger$   &    &  TPWA at 1 or 2 angles not possible\\
   &                                      &   & 12(5)3 & $I[3]\,,\check{\Sigma}[2]\,,\check{T}[3]\,,\check{P}[2]\,,\check{F}[2]$\\
   &                                      &   & 12(4)4 & $I[4]\,,\check{\Sigma}[2]\,,\check{F}[3]\,,\check{H}[3]$\\
 \hline
7 &$L=0,1,2\; (E_{0+},\ldots,M_{2+})$  & $\dagger$   &    &  TPWA at 1 or 2 angles not possible\\
& full set of 8 $S,P,D$ wave multipoles &   & 16(6)3 & $I[3]\,,\check{\Sigma}[3]\,,\check{T}[3]\,,\check{P}[3]\,,\check{F}[3]\,,\check{G}[1]$\\
   &                                    &   & 16(5)4 & $I[4]\,,\check{\Sigma}[3]\,,\check{T}[3]\,,\check{P}[3]\,,\check{F}[3]$\\
   &                                    &   & 16(4)5 & $I[5]\,,\check{\Sigma}[3]\,,\check{F}[4]\,,\check{H}[4]$\\
 \hline
\end{tabular}
\end{table}

Table~\ref{tab3} gives examples of measurement sets involving from one to a
generalized number of $4L$ multipoles using one or more angles in the TPWA. Set
1 is the trivial case and set 2 with $J=1/2$ has already been discussed in
detail. Besides the $J=1/2$ TPWA set with the minimal number of 3 observables,
but more than one angle, a solution also exists at one angle with 4
observables, which is fully equivalent to the CEA. In a set with 3 $S,P$ wave
multipoles, $E_{0+},E_{1+},M_{1-}$, 3 amplitudes are linearly independent,
e.g.\ $F_1,F_2,F_3$, and both CEA and TPWA are again equivalent. Also, with
TPWA at more than one angle, the number of observables can be reduced. Taking
the full angular distribution, a minimal set of 3 polarization observables is
already complete.

The next logical set is the full set 4 of $S$ and $P$ wave multipoles,
$E_{0+}$, $M_{1-}$, $E_{1+}$, and $M_{1+}$, but this yields a surprising
result. As already discussed, from Eqs.~(\ref{eq:CGLN}) only 3 amplitudes are
linearly independent, leading to a CEA, which is not sufficient to resolve all
4 multipoles. This is only possible by using the angular distribution of the
observables, in the minimal case by measurements at a second angle.
At this point, it is very interesting to note that solutions with only 4 observables are also
possible~\cite{tiator16}. Here we give the set of observables, ($I$, $\check{\Sigma}$, $\check{F}$,
$\check{H}$), providing a solution with no recoil measurements required. This is a very surprising
result, as it goes beyond the studies of Omelaenko~\cite{omel,wunder}, where unique solutions were
found only with 5 or more observables.

A simple set with 4 multipoles and 4 independent amplitudes is set 5 of
Table~\ref{tab3} with $E_{0+}$, $E_{1+}$, $M_{1-}$, and $E_{2-}$. In this case
also $F_4=-3E_{2-}$ is finite. For this set an equivalent set of 8 observables
yields unique solutions for a CEA, with 4 transversity amplitudes, and a TPWA, at a single
angle, with 4 multipoles. However, taking into account the angular distribution,
the number of necessary observables can be reduced to only 3, $I$,
$\check{\Sigma}$ and $\check{T}$, where no recoil measurement would
be needed.

Truncating the multipole series in the total spin $J$ (instead of angular momentum $L$) leads to
set 6 with limit $J=3/2$. This set contains 6 multipoles, and a CEA at one angle is certainly no
longer sufficient to determine all of them. The last set 7 of Table~\ref{tab3} is the full set of 8
multipoles for $L=2$ and can be generalized for any higher $L$. Here also the CEA is no longer
related to the TPWA.

The Omelaenko method~\cite{omel,wunder} can be applied to any given $L$. This
method proves, in general, a unique solution is possible with 5 observables
measured over the full angular range, i.e. at enough angles to determine the
$\cos\,\theta$ or alternatively the Legendre coefficients. These are 4
observables from group $S$, the unpolarized cross section and the 3 single-spin
polarizations, plus one more double polarization observable from any other
group, except $\check{E}$ and $\check{H}$. The 5th observable is needed to
resolve, first of all, the double ambiguity. The new solution with only 4
observables~\cite{tiator16}, which was found to provide a solution for set 4,
has been found to solve set 7 as well, and can most likely be generalized for
any higher $L$.

However, as discussed in Refs.~\cite{omel,wunder}, an increasing number of
$4^{2L}$ accidental ambiguities can occur, which leads to enormous numerical
problems for $L>2$. This problem can partly be solved by extending the set of
observables. However, the accidental ambiguities depend on the dynamics of the
underlying models and the physics involved, and unique solutions cannot be
guaranteed in many cases, so elaborate numerical methods need to be applied.
Since experimental data contain sizable statistical errors, and in most cases
also systematic errors, a unique solution by this method will become
increasingly difficult
for larger $L$. Therefore, in practice, higher partial waves have to be fixed
by models or if possible by theoretical constraints such as unitarity,
analyticity and fixed-t dispersion relations.

Instead of doing model applications, the results of Table~\ref{tab3} have been obtained in a more
general numerical simulation procedure. The underlying multipoles numbering from two to eight were
randomly chosen as complex numbers with integer values for their real and imaginary parts. From
these multipoles, all observables and their coefficients were calculated and the inverse solution
was searched with numerical minimization techniques using random search with the help of
Mathematica. Sets 1 to 6 were quickly obtained but set 7, for $L=2$, required a significant
increase in computation time. Nevertheless, the uniqueness of the solution in terms of the squared
numerical deviation is found to be of order $10^{-20}$.

\section{Conclusions}

We have explored the CEA and TPWA, applying a number of approaches, in order to compare the
information required for a complete solution. The connection is seen most easily in the first
non-trivial case, $J=1/2$, involving the interference of two multipoles or helicity/transversity
amplitudes. The reduced number of observable types for a TPWA is compensated by additional angular
measurements. From a physical standpoint, the appearance of $\theta$-dependent factors in
Eq.~(\ref{eq:B1JOneHalf}) is due to rotational symmetry, as contained in the rotation matrices used
to construct the helicity amplitudes~\cite{Walker,JW}.

This matching of information required to determine either the multipoles or helicity/transversity
amplitudes holds only when the number of independent helicity/transversity
amplitudes, for a CEA, is the same as the number of multipoles used in their construction.
The number of angular measurements for a TPWA
grows with increasing angular momentum cutoff, as described
in Refs.~\cite{omel,wunder}. With greater than four multipole amplitudes included,
the TPWA and CEA problems are fundamentally
different and the information required for a solution is not comparable.

Our pedagogical study of the simple $J=1/2$ case, generalized to higher angular-momentum cutoffs,
has revealed further solutions of the TPWA problem addressed by Omelaenko~\cite{omel}, which
require only 4 well selected polarization observables. These will be examined in detail in a future
publication~\cite{tiator16}.

\begin{acknowledgments}
The work of HH, MD, and RW was supported in part by the U.S. Department of
Energy Grant DE-SC0016582. M.D. is also supported through the NSF PIF grant No.
1415459, an NSF CAREER grant No. PHY-1452055 and the U.S. Department of Energy,
Office of Science, Office of Nuclear Physics under contract DE-AC05-06OR23177.
The work of LT and YW was supported by the Deutsche Forschungsgemeinschaft (SFB
1044 and SFB/TR16).
\end{acknowledgments}




\begin{thebibliography}{99}

\bibitem{nn} See, for example, J.~Ball {\it et al.}, Eur.\ Phys.\ J.\
    C~\textbf{5}, 57 (1998), and references therein.

\bibitem{chiang} W.-T. Chiang and F. Tabakin, Phys.\ Rev.\ C~{\bf 55}, 2054
    (1997).

\bibitem{bds} I.S. Barker, A. Donnachie, and J.K. Storrow, Nucl.\ Phys.\ B~{\bf
    95}, 347 (1975).

\bibitem{vrancx} T. Vrancx, J. Ryckebusch, T. Van Cuyck and P. Vancraeyveld,
    Phys.\ Rev.\ C~{\bf 87}, 055205 (2013).

\bibitem{nys} J. Nys, T. Vrancx and J. Ryckebusch, J. Phys.\ G~{\bf 42}, 034016
    (2015).

\bibitem{sandorfi} A.M.~Sandorfi, S. Hoblit, H. Kamano, T.-S. H. Lee, J.\
    Phys.\ G~\textbf{38}, 053001 (2011).

\bibitem{wunder} Y.~Wunderlich, R. Beck, and L.~Tiator, Phys.\ Rev.\
    C~\textbf{89}, 055203 (2014).

\bibitem{rw1} R.L. Workman, Phys.\ Rev.\ C~\textbf{83}, 035201 (2011).

\bibitem{omel} A.S. Omelaenko, Sov.\ J. Nucl.\ Phys.~{\bf 34}, 406 (1981).

\bibitem{grushin} V.F. Grushin, A.A. Shikanyan, E.M. Leikin, and A. Ya.
    Rotvain, Yad.\ Fiz.~{\bf 38}, 1448 (1983); V.F. Grushin, in {\it
    Photoproduction of Pions on Nucleons and Nuclei}, edited by A.A. Komar
    (Nova Science, New York, 1989), p.~1ff.

\bibitem{Walker} R.L. Walker, Phys.\ Rev.~\textbf{182}, 1729 (1969).

\bibitem{vpi90} R.A. Arndt, R.L. Workman, Z.~Li, L.D.~Roper, Phys.\ Rev.\
    C~\textbf{42}, 1853 (1990); 1864 (1990).

\bibitem{Barker75} I.S.~Barker, A.~Donnachie, and J.K.~Storrow, Nucl.\ Phys.\
    B~\textbf{79}, 347 (1975).

\bibitem{KW} G. Keaton and R. Workman, Phys.\ Rev.\ C~\textbf{53}, 1434 (1996).

\bibitem{tiator16} L. Tiator \textit{et al.}, in preparation.

\bibitem{JW} M. Jacob and G.C.~Wick, Ann.\ Phys.~\textbf{7}, 404 (1959).


\end{thebibliography}
\end{document}